\documentclass[letter, twocolumn]{jpsj3}
\usepackage{txfonts}
\usepackage[dvips]{color} 

\graphicspath{{fig/}} 

\begin{document} 
 
\title{ 
Field Evolution of the 
Fulde-Ferrell-Larkin-Ovchinnikov
State in a Superconductor
with Strong Pauli Effects 
} 

\date{} 
 
\author{Kenta M. \surname{Suzuki},
\name{Yasumasa \surname{Tsutsumi}}, 
\name{Noriyuki \surname{Nakai}$^{1}$}, 
\name{Masanori \surname{Ichioka}}, and
\name{Kazushige \surname{Machida}}}
\inst{
Department of Physics, Okayama University, 
\address{Okayama 700-8530, Japan} \\ 
$^{1}$Nanoscience and Nanotechnology Research Center (N2RC), Osaka Prefecture University, 
\address{1-2 Gakuen-cho, Sakai 599-8570, Japan}
}
\abst{
The Fulde-Ferrell-Larkin-Ovchinnikov (FFLO) phase in the vortex 
lattice state 
is quantitatively studied using the self-consistent Eilenberger theory 
in three-dimensional (3D) space. 
We estimate free energy to determine 
the FFLO phase diagram in the $H$-$T$ plane and stable FFLO wave number 
in the isotropic system with the 3D Fermi sphere and {\it s}-wave pairing.
To facilitate the experimental identification of the FFLO state, 
we investigate the field evolution
of NMR spectra and flux line lattice form factors 
obtained in neutron scattering
in the FFLO vortex states. 
Possible 
applications of our results to experimental data on CeCoIn$_5$
are mentioned.
}
\kword{heavy-fermion superconductors, Fulde-Ferrell-Larkin-Ovchinnikov state, 
vortex state, quasi-classical Eilenberger theory}
%

\maketitle 
There has been much attention focused on discovering and hunting 
exotic superconducting states. 
Among them, in the singlet pairing category, the Fulde-Ferrell-Larkin-Ovchinnikov
(FFLO) state has been one of the most elusive examples since its theoretical
prediction in 1964\cite{FF,LO}.
In the FFLO states, the superconducting order parameter 
exhibits a spatial modulation\cite{matsuda}.
The population imbalance is brought about either 
by the application of an external field in the charged electron case 
through the Zeeman effect 
or by the preparation 
of up and down species
 in cold neutral atom gases \cite{mizushima,machida,gases}.
Under an applied field,
the FFLO state is expected to be the one most likely 
to emerge in the low-temperature ({\it T}) high-field ({\it H}) region.
So far there is no direct evidence to prove the existence of the FFLO state 
in either the charged or neutral system.
Thus the FFLO state still remains elusive.
In the charged system, the orbital depairing effect due to electron diamagnetic motion in a magnetic field cannot be ignored since it may affect the stability of the FFLO state.

On the other hand, 
there has been no
microscopic calculation of the field evolution of the FFLO state that fully takes account of  
the orbital depairing effect, namely, the vortex effect beyond the Ginzburg-Landau (GL)
framework valid near the upper critical field ($H_{c2})$\cite{ikeda}. 
In particular, the Larkin-Ovchinnikov (LO) state with a periodically modulated amplitude of the order parameter 
is highly difficult to describe owing to the solitonic spatial variation with infinitely 
many higher harmonics in general \cite{nakanishi}. 
In CeCoIn$_5$, it is suggested that the LO state is realized rather than the 
Fulde-Ferrell (FF) state in which only the phase is modulated in the order parameter.\cite{matsuda}
In the LO state, there are two possible modulation directions with respect to the applied magnetic field:
longitudinal and  transverse. 
In this letter, we consider the longitudinal LO state in the vortex lattice.
Hereafter, the longitudinal LO vortex state will simply be called the FFLO state.
To investigate the stability of the FFLO state against a magnetic field, we solve the microscopic Eilenberger equations self-consistently in the three-dimensional (3D) space composed of the in-plane vortex lattice and the longitudinal FFLO modulation, taking into account the orbital and Pauli-paramagnetic depairings on an equal footing. 

Furthermore, to provide fundamental theoretical information on physical quantities 
in the FFLO state, we examine the effects of the FFLO modulation 
on the nuclear magnetic resonance (NMR)~\cite{curro,vesna,kumagai1,kumagai2}
and flux line lattice (FLL) form factors obtained 
in a small-angle neutron scattering (SANS)~\cite{kenzelmann,morten} 
experiment. 
It will become apparent that these two methods can provide direct and crucial evidence of the FFLO state,
among the variety of other experiments\cite{matsuda,bianchi}. 
We will discuss anomalous behaviors in the corresponding experimental data 
on the heavy Fermion superconductor CeCoIn$_5$, the high-field and low-temperature 
superconducting phase of which is regarded to be a realization of the FFLO state. 
Thus the main purpose of this letter is to demonstrate
the NMR spectrum and FLL form factors through the theoretical study. 

Our basic strategy is to provide 
$H$-dependent properties 
of the FFLO states for the 3D Fermi sphere and $s$-wave pairing.
The corresponding 3D calculation for the FF state \cite{takada} 
and the full self-consistent analytical theory 
for a quasi-1D case\cite{nakanishi} have been performed 
previously in the Pauli limiting case 
without vortices.
Here, we extend those calculations to take account of 
vortex effects. 
Before discussing the anomalous behavior of the FFLO states in CeCoIn$_5$, 
it is necessary to clarify the quantitative properties of 
the FFLO state in a typical example of the 3D Fermi sphere. 


We calculate the spatial structure of the vortex lattice state
using the quasi-classical Eilenberger theory in the clean 
limit valid for $k_{\rm F}\xi \gg 1$ ($k_{\rm F}$ is the Fermi wave number and 
$\xi$ is the superconducting coherence length)~\cite{ichiokaS,hasegawa}.
The Pauli paramagnetic effects are included 
through the Zeeman term $\mu_{\rm B}B({\bf r})$, 
where $B({\bf r})$ is the flux density of an internal field and 
$\mu_{\rm B}$ is a renormalized Bohr 
magneton.
The quasi-classical Green's functions
$g( \omega_n +{\rm i} {\mu} B, {\bf k},{\bf r})$, 
$f( \omega_n +{\rm i} {\mu} B, {\bf k},{\bf r})$, and 
$f^\dagger( \omega_n +{\rm i} {\mu} B, {\bf k},{\bf r})$  
are calculated in the vortex lattice state  
by the Eilenberger equations~\cite{ichioka,ichiokaFFLO} 
\begin{eqnarray} &&
\left\{ \omega_n +{\rm i}{\mu}B 
+\tilde{\bf v} \cdot\left(\nabla+{\rm i}{\bf A} \right)\right\} f
=\Delta g, 
\nonumber 
\\ && 
\left\{ \omega_n +{\rm i}{\mu}B 
-\tilde{\bf v} \cdot\left( \nabla-{\rm i}{\bf A} \right)\right\} f^\dagger
=\Delta^\ast g  , \quad 
\label{eq:Eil}
\end{eqnarray} 
where $g=(1-ff^\dagger)^{1/2}$, ${\rm Re} g > 0$, 
$\tilde{\bf v}={\bf v}/v_{{\rm F}0}$, 
and the Pauli parameter ${\mu}=\mu_{\rm B} B_0/\pi k_{\rm B}T_{\rm c}$. 
${\bf k}$
 is the relative momentum of the Cooper pair, 
and ${\bf r}$ is the center-of-mass coordinate of the pair. 
${\bf v}$ is the Fermi velocity and
$v_{\rm F0}=\langle v^2 \rangle_{\bf k}^{1/2}$ 
where $\langle \cdots \rangle_{\bf k}$ indicates the Fermi surface average. 
We assume 
a magnetic field is applied to the z-axis.
The Eilenberger units of $R_0$ for lengths and $B_0$ for a magnetic field 
are used\cite{ichioka,ichiokaFFLO}.
The order parameter $\Delta$ and the Matsubara frequency $\omega_n$ 
are normalized in units of $\pi k_{\rm B} T_{\rm c}$.

For self-consistent conditions, 
the order parameter is calculated by 
\begin{eqnarray}
\Delta({\bf r})
= g_0N_0 T \sum_{0 < \omega_n \le \omega_{\rm cut}} 
 \left\langle 
    f +{f^\dagger}^\ast 
\right\rangle_{\bf k} 
\label{eq:scD} 
\end{eqnarray} 
with 
$(g_0N_0)^{-1}=  \ln T +2 T
        \sum_{0 < \omega_n \le \omega_{\rm cut}}\omega_n^{-1} $. 
We use $\omega_{\rm cut}=20 k_{\rm B}T_{\rm c}$.
${\bf B}=\nabla\times{\bf A}$ 
is self-consistently determined by 
\begin{eqnarray}
\nabla\times \left( \nabla \times {\bf A} \right) 
=\nabla\times {\bf M}_{\rm para}({\bf r})
-\frac{2T}{{{\kappa}}^2}  \sum_{0 < \omega_n} 
 \left\langle \tilde{\bf v} 
         {\rm Im}~g  
 \right\rangle_{\bf k}, 
\label{eq:scH} 
\end{eqnarray} 
where we consider both the diamagnetic contribution of 
supercurrent in the last term 
and the contribution of the paramagnetic moment 
${\bf M}_{\rm para}({\bf r})=(0,0,M_{\rm para}({\bf r}))$ 
with 
\begin{eqnarray}
M_{\rm para}({\bf r})
=M_0 \left( 
\frac{B({\bf r})}{H} 
- \frac{2T}{{\mu} H }  
\sum_{0 < \omega_n}  \left\langle {\rm Im}~g 
 \right\rangle_{\bf k}
\right) . 
\label{eq:scM} 
\end{eqnarray} 
The normal state paramagnetic moment 
$M_0 = ({{\mu}}/{{\kappa}})^2 H $,   
${\kappa}=B_0/\pi k_{\rm B}T_{\rm c}\sqrt{8\pi N_0}$  and 
$N_0$ is the density of states at the Fermi energy in the normal state. 
We set the GL parameter $\kappa$ to be $102$.
We solve eq. (\ref{eq:Eil}) and eqs. (\ref{eq:scD})-(\ref{eq:scM})
alternately, and obtain self-consistent solutions, 
as in previous works~\cite{ichioka,ichiokaFFLO},
under a given unit cell of the triangular vortex lattice. 
For the FFLO state, 
$\Delta({\bf r})$ has a periodic oscillation with the period $L$ 
in addition to the vortex lattice structure. 
As the unit cell size of the vortex lattice is determined by $H=\langle {\bf B}\rangle_{\bf r}$, 
we can estimate the $H$-dependence of the FFLO state in our calculation of the vortex lattice.  
Throughout this paper we use $\mu=5$
as a representative case of the strong Pauli paramagnetic effect.

\begin{figure}[tb]
\centering 
\includegraphics[width=5.5cm]{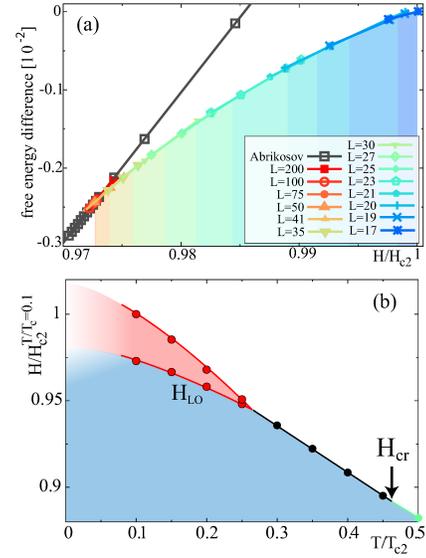} 
 \caption{ 
(Color online)  
(a) Free energy differences $F$ from the normal state 
for the FFLO state with different wave numbers $L$ and the Abrikosov state, 
as a function of $H$. $T = 0.1T_c$. 
In the normal state, $F=0$. 
(b) Phase diagram for the FFLO state in the $H$-$T$ plane for the 3D Fermi sphere
and the $s$-wave pairing. $\mu$= 5. 
$H_{c2}$ is the first order at $H>H_{\rm cr}$. 
Lines are guides for the eye. 
} 
\label{fig:phase-diagram} 
\end{figure} 

\begin{figure}[tb] 
\centering
\includegraphics[width=5.5cm]{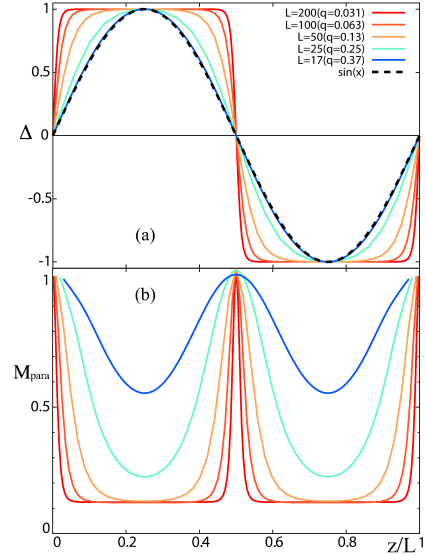} 
 
\caption{ 
(Color online)  
Spatial variations of (a) the order parameter $\Delta({\bf r})$ 
and (b) the paramagnetic moment $M_{\rm para}({\bf r})$ 
for several values of $L$ along the field direction outside of vortices.
These are normalized by its length and maximum values.
$T/T_c = 0.1$ and $\mu=5$.
} 
\label{fig:pair-potential} 
\end{figure} 

The Gibbs free energies are calculated from self-consistent solutions 
using eq. (9) in ref. \citen{hiragi} 
for the FFLO state with various FFLO wavelengths $L$. 
We compare them to identify the most stable state under a given $H$ and $T$. 
Figure \ref{fig:phase-diagram}(a) exhibits the
resulting successive changes at $T/T_c=0.1$.
It is seen that,  starting from $H=H_{c2}$ where the FFLO state with the
shortest wavelength $L=17$ is stabilized, 
$L$ becomes longer as $H$ decreases.
 Eventually the free energy of the FFLO state becomes comparable to that 
 of the conventional Abrikosov state, 
where the FFLO modulation along the field direction is absent.
 The envelope of the free energies of the FFLO state approaches that
 of the Abrikosov state, 
such that the two curves seem to merge tangentially, namely,
 at the meeting point, tangents of the two curves coincide with 
 each other. 
While our calculations are for discretized $L$, 
even these results suggest 
(1) a second-order-like transition between 
the FFLO state and the Abrikosov vortex state\cite{ikeda} and (2) the continuous
$L$ change of the FFLO state as a function of $H$, 
which are similar to the results of previous analytic FFLO theory\cite{nakanishi}.

In Fig. \ref{fig:phase-diagram}(b),  
we show the phase diagram in the $H$-$T$ plane. 
This is obtained by repeating the FFLO calculations as a function of $H$ 
at different temperatures, 
$T/T_c$=0.1, 0.15, 0.2, and 0.25. 
$H_{LO}$ is the transition field from the Abrikosov vortex state to 
the FFLO state
and the transition at $H_{c2}$ to the normal phase is the first order above $H_{cr}$.
The FFLO region in the $H$-$T$ plane is given by $H_{\rm LO}/H_{c2}=0.973$
for the present $\mu=5$ at $T/T_c$=0.1.
The value of $H_{\rm LO}/H_{c2}$ depends on the $\mu$ value, namely, 
$H_{\rm LO}/H_{c2}=0.991$ for $\mu=2$ at $T/T_c$=0.1. 
Even in this strong paramagnetic case of $\mu=5$, 
the FFLO phase appears only near $H_{c2}$, and 
$H_{\rm LO}$ increases on lowering $T$  
in this typical example of an isotropic Fermi sphere. 

Figure \ref{fig:pair-potential} displays 
normalized waveforms of (a) the order parameter $\Delta({\bf r})$ 
and (b) the paramagnetic moment $M_{\rm para}({\bf r})$ in the FFLO state 
along the field direction outside of vortices. 
It is seen that a simple sinusoidal modulation waveform continuously
deforms into an anti-phase kink form, or solitonic waveform as 
$H$ approaches the $H_{\rm LO}$ line where $L$ diverges.
In other words, near the $H_{\rm LO}$ boundary, 
the sign change or $\pi$ phase shift of the order parameter occurs sharply, 
meaning that the excess electrons and $M_{\rm para}({\bf r})$ 
are confined in a narrow spatial region along the kink position. 

The FFLO nodal kink forms a sheet of paramagnetic moments 
perpendicular to the field. 
On the other hand, 
along the vortex lines, 
enhanced $M_{\rm para}({\bf r})$ 
at the vortex core is decreased at the intersection with FFLO kink 
plane
(see Fig. 2 in ref. \citen{ichiokaFFLO} and Fig.1 in ref. \citen{mizushima2}).
There, the zero-energy peak states of quasiparticles are absent,  
because of the 2$\pi$ phase shift of the order parameter,  coming from the kink and from the vortex. 
The paramagnetic moment  becomes strongly confined to the
kink position as $H$ approaches $H_{\rm LO}$ from above.
As will be seen later, these three-dimensional FFLO spatial structures
can be probed by SANS experiments or 
NMR experiments.

\begin{figure}[tb]  
\centering
 \includegraphics[width=5.5cm]{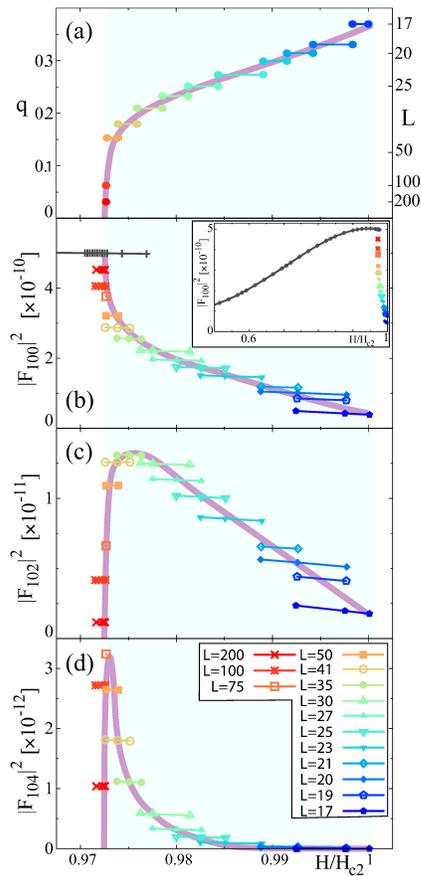}
\caption{ 
(Color online)  
Field evolutions of various quantities at $T/T_c=0.1$
and $\mu=5$. 
(a) FFLO wave number $q=2 \pi/L$.
(b) Form factor $|F_{100}|^2$. Inset shows the overall variation.
(c) Form factor $|F_{102}|^2$.
(d) Form factor $|F_{104}|^2$.
} 
\label{fig:SANS} 
\end{figure} 

As shown in Fig. \ref{fig:SANS}(a), 
the FFLO wave number $q=2 \pi/L$ of the stable FFLO state continuously varies with $H$.
Starting with $q=0$ at $H=H_{\rm LO}$, \textit{q} increases sharply. 
Hence the antiphase solitonic waveform quickly changes 
into a sinusoidal one on increasing $H$ 
(see also Fig. \ref{fig:pair-potential}). 
This behavior is similar to that seen 
in the exact solution (see Fig. 9 in ref. \citen{nakanishi}), 
implying that the FFLO physics along the parallel direction
exemplified here is common and universal.

\begin{figure}
\centering
\includegraphics[width=6.cm]{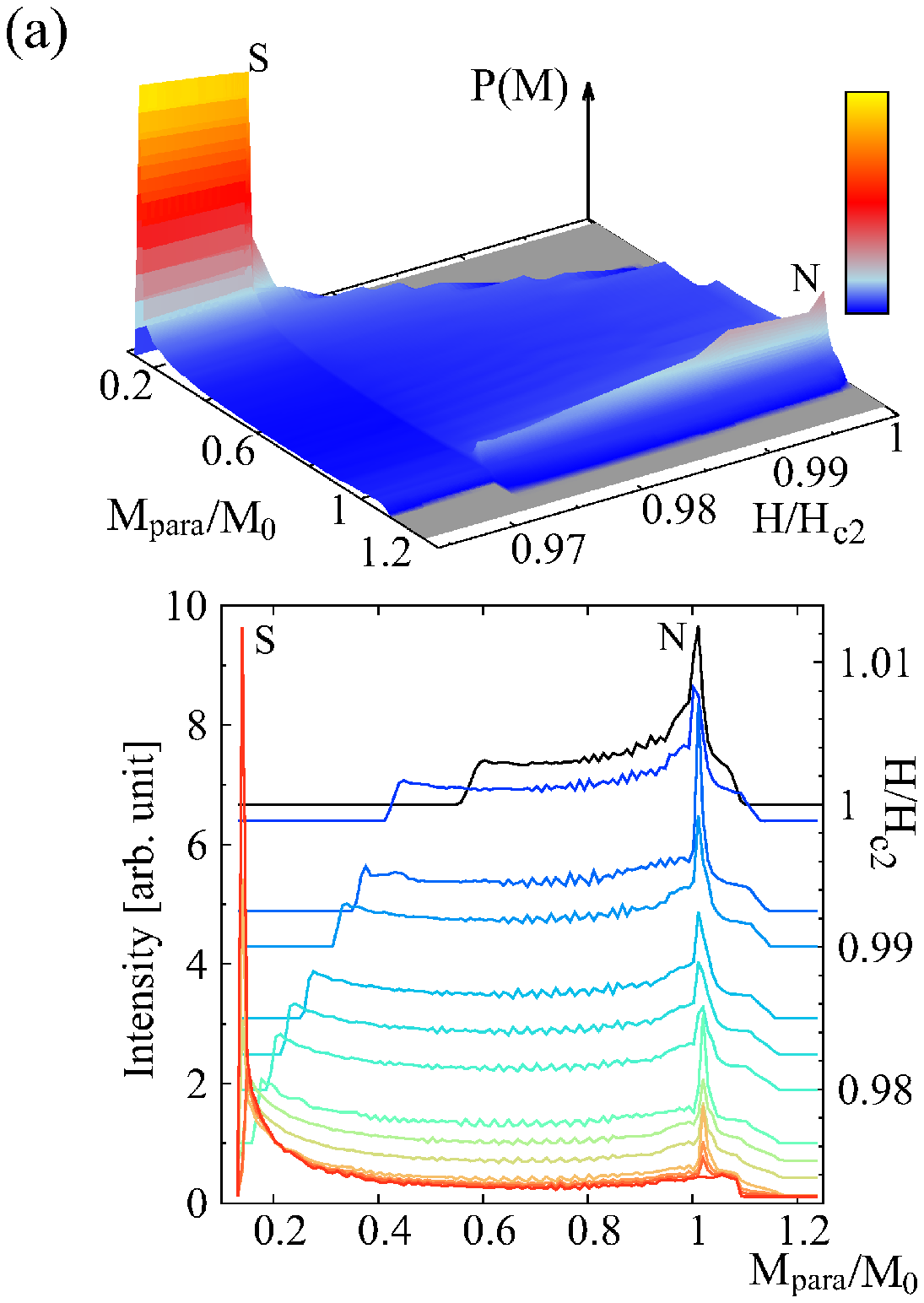}%
\\
\vspace{0.5cm}
\includegraphics[width=6.cm]{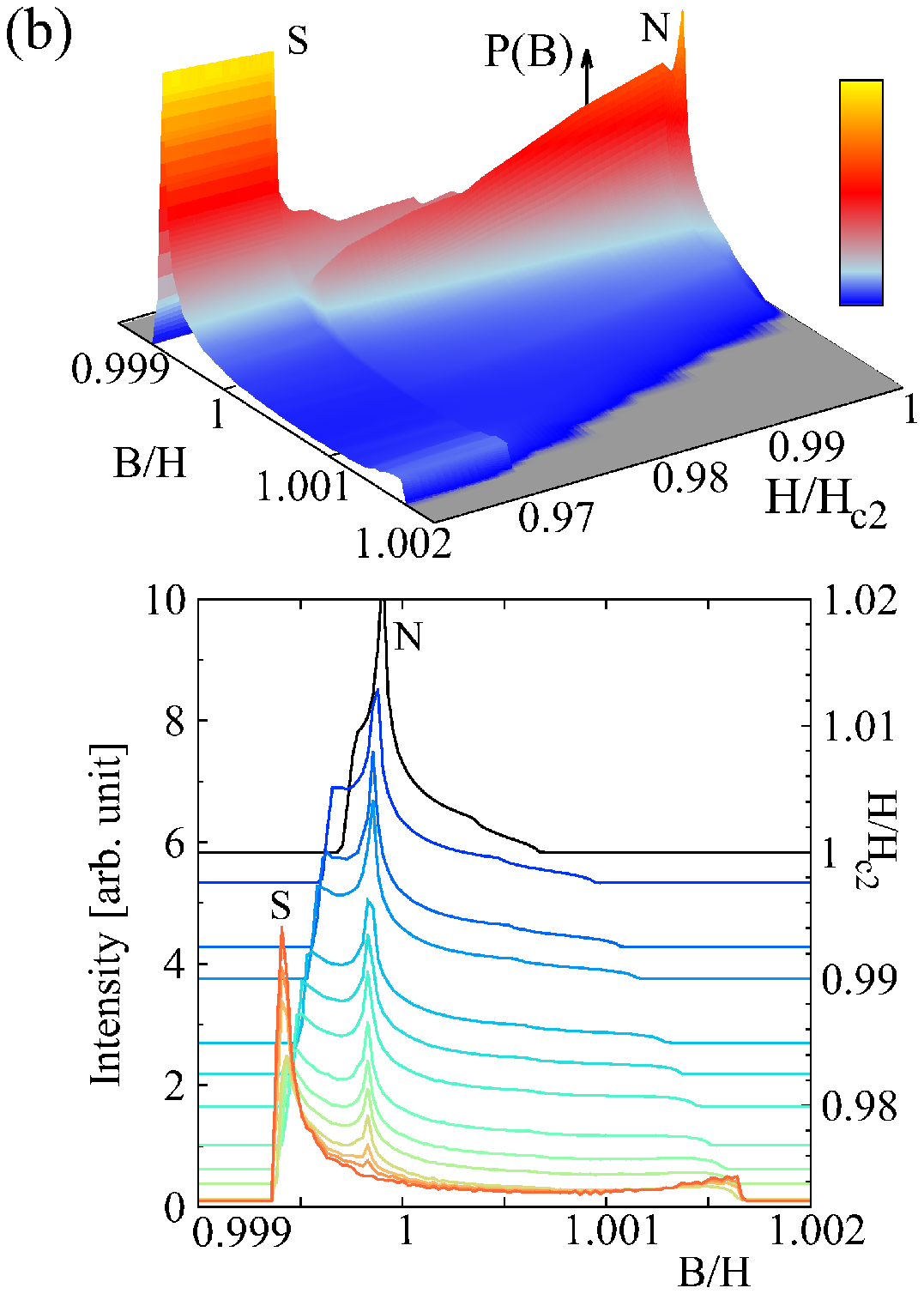}%
\caption{
(Color online) 
NMR spectra in the FFLO state: 
(a) paramagnetic moment distribution $P(M)$ 
(b) and internal field distribution $P(B)$. 
$\mu=5$ and $T/T_c=0.1$.
Upper panels show $H$-evolution of spectra 
in stereographic view. 
Lower panels show spectra at some values of $H$. 
Horizontal baselines for each spectrum are shifted by $H/H_{\rm c2}$, 
which is indicated on the right axis. 
}
\label{fig:NMR}
\end{figure}

The FLL form factor is an important quantity that can be directly measured in a
SANS experiment. 
The form factors $F_{hkl}$ are Fourier components 
of an internal field ${\bf B}({\bf r})$ 
in our calculation.~\cite{ichiokaFFLO} 
The fundamental Bragg spots $F_{100}$ for the vortex lattice 
are shown in Fig. \ref{fig:SANS}(b) as a function of $H$.
The spot $|F_{100}|^2$ increases in the Abrikosov state,  
because $M_{\rm para}({\bf r})$ accumulates at the vortex core
to increase $B({\bf r})$ locally, as seen in the inset 
of Fig. \ref{fig:SANS}(b). 
This feature has already been shown theoretically\cite{ichioka} 
and observed in various paramagnetically enhanced superconductors,
such as TmNi$_2$B$_2$C\cite{Tm} and  CeCoIn$_5$\cite{morten}.
As shown in Fig. \ref{fig:SANS}(b), 
the intensity of $|F_{100}|^2$ suddenly decreases upon entering the
FFLO phase and continues to drop quickly, almost exponentially.
(Note the $T=50$ mk data in Fig. 1 of ref. \citen{white}).
This is because $M_{\rm para}({\bf r})$ 
is not enhanced at the vortex core on the FFLO nodal plane
%
(see Fig. 5(b) in ref. \citen{ichiokaFFLO}).
This contribution decreases $|F_{100}|^2$, which is the average along
the $z$-axis.

In addition to the usual Bragg spots $F_{100}$
associated with the vortex lattice, 
the observation of extra spots $F_{10n}$ ($n=$2,4, ...) 
is crucial to prove the existence of the FFLO phase. 
In Fig. \ref{fig:SANS}(c), 
we show $|F_{102}|^2$, which is the new superspot associated with 
the FFLO modulation along the field direction. 
It rises quickly at $H=H_{\rm LO}$.
After attaining a maximum in the middle of the FFLO phase, 
it slowly decreases towards $H_{c2}$.
Thus the best chance to observe it is in the middle field region 
inside the FFLO phase.
The relative intensity $|F_{102}|^2/|F_{100}|^2= 1/10 \sim 1/20$.
Therefore it is quite possible to detect it 
because $|F_{100}|^2$ is enhanced by the Pauli effect even near $H_{c2}$.
The higher order spot $|F_{104}|^2$ is also shown in Fig. \ref{fig:SANS}(d). 
It takes a maximum just near $H_{\rm LO}$. 
Since the magnitude of $|F_{104}|^2$ is further reduced 
and is one order of magnitude smaller than $|F_{102}|^2$,
it might be difficult to detect it.

The NMR spectrum is also crucial to identify the FFLO state.
By choosing probed nuclei that have different hyperfine coupling constants, 
we can effectively pick up the selective field distributions\cite{ichiokaFFLO}.  
When the hyperfine coupling is sufficiently strong, 
the paramagnetic distribution $M_{\rm para}({\bf r})$ is probed in NMR experiments.
In contrast, in the weak coupling case, the magnetic induction $B({\bf r})$ 
in the whole system is detected by NMR.
In the mixed state of ordinary superconductors, it 
yields the so-called Redfield pattern.

Here, we examine the field evolution of the NMR spectra for both strong and weak
hyperfine coupling cases. For the former (latter), we evaluate the distribution $P(M)$
[$P(B)$] using the stable FFLO state at each field. 
These are given by 
\begin{eqnarray} 
P(M)=\int\left( M-M_{\rm para}({\bf r}) \right) {\rm d}{\bf r}, \ 
P(B)=\int\left( B-B({\bf r}) \right) {\rm d}{\bf r}, 
\end{eqnarray} 
i.e., volume counting for each $M$ and $B$. 
Figure \ref{fig:NMR} shows the spectral evolutions of these distributions for two cases.
In Fig. \ref{fig:NMR}(a), $P(M)$ is displayed.
Since in the Abrikosov state the paramagnetic moment is confined exclusively 
to the vortex cores, a single main peak appears at the saddle point (S) position in the NMR spectrum.
In the FFLO phase, $M_{\rm para}({\bf r})$,  
which comes from excess electrons, accumulates at the normal state ($N$) position. 
The peak at the N-position becomes dominant towards $H_{c2}$,  
because an increasing excess of unpaired quasi-particles 
appear at the FFLO nodal sheets. 
It is noted that near $H_{\rm LO}$, two peaks appear simultaneously in 
the NMR spectrum.
This double peak structure was observed in the In(2) site of the NMR spectrum 
by Kumagai {\it et al.}\cite{kumagai2} for CeCoIn$_5$.
The appearance of the double peaks at S- and N-positions is unambiguous
evidence of the FFLO state.

It is also important to observe the characteristic change of $P(B)$ for
the weak hyperfine case, exemplified by In(1) in CeCoIn$_5$.
Here, rather unexpectedly, the double peak structure can be seen in
Fig. \ref{fig:NMR}(b) near $H_{\rm LO}$, 
beyond which the $N$ peak dominates the spectrum. 
The N-position is near the S-position in $P(B)$, compared with $P(M)$. 
In the lower field of the Abrikosov state, the usual Redfield pattern 
is reproduced, as seen from Fig. \ref{fig:NMR}(b).

We touch upon the recent NMR experiment on CeCoIn$_5$ \cite{kumagai2}. 
The observed double peak structure of In(2a) for $H \parallel c$
and $H\parallel ab$ is markedly similar to our result in Fig. \ref{fig:NMR}(a)
(see the spectral evolution in Fig.2 of ref. \citen{kumagai2}).
The proposed phase diagram of FFLO for $H\parallel c$ is also 
similar to our Fig. \ref{fig:phase-diagram}(b) 
where 
$H_{\rm LO}/H_{c2}\sim 0.975$ 
for $\mu=5$
compared with $H_{\rm LO}/H_{c2}=4.7T/4.95T\sim 0.95$ at zero temperature
for $H\parallel c$.\cite{kumagai1} As mentioned previously, the value of $H_{\rm LO}/H_{c2}$
depends on $\mu$, but the shape of the FFLO phase diagram
is hardly changed by the value of $\mu$. For $H\parallel ab$, the proposed 
phase diagram is  modified because of the presence of the existing
 SDW\cite{suzuki}.

We also calculated the FFLO structure in $d$-wave pairing or 
the quasi-two-dimensional Fermi surface 
when a magnetic field is applied to the \textit{z}-axis. 
These situations do not qualitatively change 
the results shown 
in Figs. \ref{fig:pair-potential}-\ref{fig:NMR}. 
Careful estimation of the free energy to determine the stable 
$L$ remains for future work.

In conclusion, 
we quantitatively explored the field evolution of the FFLO state 
for typical examples of a Fermi sphere and $s$-wave pairing, 
by self-consistently solving the microscopic Eilenberger equations 
in the 3D space of the vortex lattice and the FFLO modulation along the field direction. 
To facilitate the identification of the FFLO state through experiments,
we estimate the NMR spectrum and FLL form factors 
as a function of the magnetic field in the FFLO vortex states.

  
The authors are grateful for insightful discussions with  
M. Kenzelmann, S. Gerber, J. S. White, J. L. Gavilano, T. Sakakibara, E. M. Forgan,
K. Kumagai, R. Ikeda, Y. Matsuda, and M. R. Eskildsen.


\end{document}